\documentclass[
 aip,amssymb,jcp,
reprint,floatfix,
]{revtex4-2}
\usepackage{graphicx}
\usepackage{color}
\usepackage{hyperref}
\usepackage{float}
\usepackage{array, multirow}
\usepackage{amsmath}
\usepackage{newtxtext}
\usepackage{booktabs}
\usepackage{xcolor}
\usepackage{soul}
\usepackage{url}
\usepackage{gensymb}
\usepackage[T1]{fontenc}
\usepackage[caption=false]{subfig}

\hypersetup{colorlinks,
    linkcolor={blue!75!black!80!yellow},
    citecolor={blue!75!black!80!yellow},
    urlcolor={blue!75!black!80!yellow}
    }

\begin{document}

\title{Orbital Entanglement and The Double $d$-Shell Effect in Binary Transition Metal Molecules}

\author{Julianne S. Lampert}
\affiliation{Department of Chemistry, Washington University in St. Louis, St. Louis, MO 61630 USA}

\author{Timothy J. Krogmeier}
\affiliation{Department of Chemistry, Washington University in St. Louis, St. Louis, MO 61630 USA}
\affiliation{Department of Chemistry, University of Minnesota, Minneapolis, MN 55455 USA}

\author{Anthony W. Schlimgen}
\email{aws@umn.edu}
\affiliation{Department of Chemistry, Washington University in St. Louis, St. Louis, MO 61630 USA}
\affiliation{Department of Chemistry, University of Minnesota, Minneapolis, MN 55455 USA}

\author{Kade Head-Marsden}
\email{khm@umn.edu}
\affiliation{Department of Chemistry, Washington University in St. Louis, St. Louis, MO 61630 USA}
\affiliation{Department of Chemistry, University of Minnesota, Minneapolis, MN 55455 USA}

\date{\today}

\begin{abstract}
Accurate modeling of transition metal-containing compounds is of great interest due to their wide-ranging and significant applications. These systems present several challenges from an electronic structure perspective, including significant multi-reference character and many chemically-relevant orbitals. A further complication arises from the so-called double $d$-shell effect, which is known to cause a myriad of issues in the treatment of first-row transition metals with both single- and multi-reference methods. While this effect has been well documented for several decades, a comprehensive understanding of its consequences and underlying causes is still evolving. Here, we characterize the second $d$-shell effect by analyzing the information entropy of correlated wavefunctions in a periodic series of $3d$ and $4d$ transition metal molecular hydrides and oxides. These quantum information techniques provide unique insight into the nuanced electronic structure of these species, and are powerful tools for study of weak and strong correlation in the transition metal $d$ manifold.
\end{abstract}

\maketitle

\section{Introduction}
Transition metals are a vital component of contemporary chemistry, biology, and materials science owing to a rich variety of chemical properties. The electronic structure of $d$-electrons and the nature of the metal-ligand bond have fundamental consequences in quantum technologies,~\cite{Gaita-Arino:2019,Freedman2021, Fataftah:2018aa} catalysis,~\cite{Vogiatzis:2019,Gaggioli:2019,Melius:1976aa} biophysical processes,~\cite{Siegbahn:2000aa,Schilter:2016aa,Zheng:2014aa} and spectroscopy.~\cite{Merer:1989aa, Hart:2024aa, Kuhnt:2024aa} In addition, the nature of the metal $d$-orbitals allows for interesting interactions with surrounding ligand fields as well as tunability of oxidation and spin states.~\cite{Kaim:2016aa,Swart:2016aa,Moltved:2019aa, Kepp:2016aa} This broad range of reactivity and structure in turn presents a challenge to robust computational modeling of these species.~\cite{Langhoff:1988aa,Pierloot:2011,Harrison:2000aa,Jiang:2012aa, Khedkar:2021aa} 

One computational approach uses correlated wavefunction-based techniques, and seeks to describe subtle effects of electron correlation often important in transition metal complexes. In these calculations a phenomenon known as the ``double $d$-shell" or ``second $d$-shell" effect can occur, where explicitly correlating only a $3d$ set of orbitals describes the electronic structure poorly.~\cite{Fischer:1977aa, Pierloot:1995aa} The effect has been described in a variety of contexts,~\cite{Fischer:1977aa, Dunning:1980aa, Pierloot:1995aa,Sunil:1985aa, Bauschlicher:1988aa,Bauschlicher:1988ab,Chong:1986aa,Boguslawski:2012aa,Moraes:2023aa,Andersson:1992aa,Barandiaran:2013aa,Kalemos:2008aa} but important questions remain surrounding the second $d$-shell and its relationship to different correlation effects. Transition metals are particularly sensitive to the balance of static, or multi-reference, and dynamic correlation. Roughly, the former arises when multiple Slater determinants are important for a many-electron wavefunction, while the latter describes the two-electron Coulomb repulsion. Quantum mutual information is useful for understanding these many-body interactions, and providing a rough classification of static and dynamic correlation in transition metal systems.~\cite{Stein:2017aa,Boguslawski:2012aa,Boguslawski:2013aa,Nowak:2021aa,Barcza:2011ab,Stemmle:2018aa} In this context, a systematic study of the roles of static and dynamic correlation and the second $d$-shell effect is important for informing correlated calculations for transition metal chemistry. In this \emph{Article}, we characterize the double $d$-shell effect by describing the electron correlation of a series of transition metal species using quantum information entropy techniques.

Complete active-space self-consistent field (CASSCF) is the basis of many techniques for computing energetic, magnetic, and spectroscopic properties for metal complexes.~\cite{Feldt:2022aa, Reiher:2009aa, Sauri:2011aa, Phung:2018aa, Vancoillie:2011aa} \textcolor{black}{By choosing a subset of the molecular orbitals (MOs), known as the active space, CASSCF describes the total correlation in the active space, which is embedded in a mean-field of the other electrons. CASSCF is a variational approximation to the exact solution of the Schr\"odinger equation.}~\cite{Roos:1980,Siegbahn:2000aa} \textcolor{black}{As the active space approaches the full MO basis, the calculation approaches the full configuration interaction (FCI) solution. Typically, orbitals are chosen for the active space which are important for describing the chemistry of interest. For transition metals, the valence orbitals and the $d$-orbital description are particularly important.} Because the practical limit for wavefunction-based CASSCF is about 18 orbitals, finding the minimal set of MOs is essential for all but the simplest metal compounds. Several techniques including density-matrix renormalization group (DMRG),~\cite{White:1992,Chan:2002,Schollwock:2005aa} heat-bath configuration interaction,~\cite{Holmes:2016} \textcolor{black}{restricted active-space SCF (RASSCF),\cite{Olsen:1988ab, Malmqvist:1990ab}} and reduced density matrix (RDM) optimization~\cite{Mazziotti:2011, Mazziotti:2012} have been developed to extend the possible size of the active space. Unfortunately, accurate calculation of dynamic correlation, essential for quantitative prediction of structure and properties of transition metal species, is extremely sensitive to both active space size and composition.~\cite{Sand:2017, Larsson:2022, Bhowmick:2023, Chowdhury:2023, Shiozaki:2021,Roy-Chowdhury:2018aa} 

These considerations are particularly important in the context of the second $d$-shell. Considered most important for the $3d$ metals,\cite{Pierloot:1995aa} one source of the phenomenon is that the radial extent of standard basis sets for the 3$d$ orbitals is too short-ranged to describe the true radial distribution of 3$d$ electrons.~\cite{Dunning:1980aa} Additionally, the problem becomes more severe as the number of $d$-electrons increases, because the optimal size of the 3$d$ orbitals is particularly sensitive to their occupation.~\cite{Pierloot:2001aa, Botch:1981aa, Moraes:2023aa} While this may indicate a purely dynamical effect, the second $d$-shell effect can appear as a consequence of either static or dynamic correlation, and even a combination thereof.\cite{Pierloot:1995aa,Moraes:2023aa, Li-Manni:2018aa} Tools in quantum information theory are useful for understanding the balance of these effects, in particular by computing orbital entanglement entropy of electronic states. Entanglement entropy has been used to aid in active space selection, and to understand subtle orbital interactions.\cite{Stein:2017aa,Boguslawski:2012aa,Boguslawski:2013aa,Nowak:2021aa,Barcza:2011ab,Stemmle:2018aa,Stein:2019aa}

Here we combine the analysis of traditional electronic structure metrics, such as population and vibrational frequencies, with information entropy metrics to understand how electron correlation is modulated by the second $d$-shell. Our results show that traditional metrics can be augmented with entanglement analysis to understand how active space selection can bias multi-referenced wavefunctions, especially when considering dynamical correlation corrections.

\section{Computational Methods}
The importance of the second $d$-shell effect in molecules depends on the bonding environment, as well as the electronic state of the species.~\cite{Pierloot:2011,Moraes:2023aa} To characterize the second $d$-shell effect, we compare two active spaces and use the notation [$N_e, N_o$], with $N_e$ active electrons and $N_o$ active spatial orbitals. For an $nd$ transition metal, the small active spaces contain 9 orbitals for the oxides and 7 for the hydrides. The large active spaces contain the second $d$-shell, denoted by $nd'$, bringing the number of active orbitals to 14 for the oxides and 12 for the hydrides. For the large active spaces, extra virtual orbitals were sometimes included to improve the stability of orbital rotations in the CASSCF. A full description of the active spaces and their composition can be found in the Supporting Information Tables S1-S4. 

All calculations were performed with OpenMolcas \textcolor{black}{version 23.02}~\cite{Fdez.-Galvan:2019aa,Aquilante:2020aa,Li-Manni:2023aa} using the ANO-RCC-VTZP basis with scalar relativistic corrections.~\cite{Roos2004,Roos2005,widmark1990a,Douglas1974,Hess1986} The internuclear axis of the molecules lies along the $z$-direction, such that that the $d_{z^2}$ orbitals are positioned along the internuclear axis. The geometries were optimized and frequencies calculated with complete active-space second order perturbation theory (CASPT2),~\cite{Andersson1992,Andersson1990} and the mutual information was computed from DMRG wavefunctions using QCMaquis.~\cite{Keller:2015aa, Dolfi:2014aa} The frequencies are computed numerically by finite difference, so we expect some numerical noise in those results. Computed internuclear distances are available in Supplementary Information Table S5.

We use the converged final orbitals to compute the one- and two-orbital entropy using DMRGCI with four DMRG sweeps in Fiedler ordering, and a maximum bond dimension of $M=1000$.\cite{Olivares-Amaya:2015aa} The active spaces used here are small enough for exact diagonalization, so we only use the DMRG routines to easily extract the information entropy from the wavefunction. The DMRG calculations with these parameters yield energies with an error compared to the exact solutions of about $10^{-6}$ a.u. and $10^{-4}$ a.u. for the small and large active spaces, respectively.

In this work, we use the one-orbital entropy, 
\begin{equation} \label{eqn:single_orb_ent}
    s_i= -\sum_{\alpha} \omega_{\alpha;i}\ln{\omega_{\alpha;i}}, 
\end{equation}
the two-orbital entropy,
\begin{equation} \label{eqn:two_orb_ent}
    s_{i,j}= -\sum_{\alpha} \omega_{\alpha;i,j}\ln{\omega_{\alpha;i,j}}, 
\end{equation}
and the mutual information,
\begin{equation} \label{eqn:mutinfo}
    I_{i,j}= s_{i,j} - s_i - s_j,
\end{equation}
where $\omega_{\alpha;i}$ are the eigenvalues of the one-orbital RDM and $\omega_{\alpha;i,j}$ are the eigenvalues of the two-orbital RDM.~\cite{Boguslawski:2012aa} The mutual information quantifies the shared quantum information between two orbitals beyond a one-electron picture. The magnitude of the one-orbital entropy and mutual information is also used to assess the type of electron correlation involved between the two orbitals. Mutual information is typically classified according to its order of magnitude with $10^{-1}$, $10^{-2}$, and $10^{-3}$ considered large, moderate, and small, respectively. When the one-orbital entropy is large, greater than 0.5, with large mutual information the interaction is often associated with non-dynamic correlation, which is important for bond-breaking processes. Moderate one-orbital entropy, between 0.1 and 0.5, with moderate mutual information can indicate important static correlation effects. Finally, small one-orbital entropy and small mutual information signals the importance of dynamic correlation.~\cite{Boguslawski:2012aa, Boguslawski:2013aa} 

\section{Results}

We compare sets of neutral-charge early hydrides (TiH, VH, CrH), late oxides (FeO, CoO, NiO), and their corresponding 4$d$ analogues in the neutral ground state. In Table~\ref{tab:Vib_E} we show the change in lowest vibrational frequencies upon increasing the size of the active space for the hydrides and oxides at their CASPT2 equilibrium geometries.
\begin{table}[h!]
    \centering
     \begin{tabular}{@{\extracolsep{4pt}}l c c } 
        \hline \hline
        \addlinespace[2pt]
        & $\Delta E_{vib}$ (cm$^{-1}$) & $\Delta$Charge       \\
        \cline{2-2}\cline{3-3}
        \addlinespace[2pt]
TiH	&	-16.5	&	0.029	\\
ZrH	&	11.1	&	0.020	\\
\addlinespace[2pt]
\hline
 \addlinespace[2pt]
VH	&	-17.4	&	0.019	\\
NbH	&	42.8	&	-0.004	\\
\addlinespace[2pt]
\hline
 \addlinespace[2pt]
CrH	&	1,150	&	0.062	\\
MoH	&	-71.9	&	-0.001	\\
\addlinespace[2pt]
\hline\hline
 \addlinespace[2pt]
FeO	&	-65.0	&	0.193	\\
RuO	&	-41.3	&	0.076	\\
\addlinespace[2pt]
\hline
 \addlinespace[2pt]
CoO	&	-13.6	&	0.161	\\
RhO	&	9.62	&	0.076	\\
\addlinespace[2pt]
\hline
 \addlinespace[2pt]
NiO	&	186	&	0.223	\\
PdO	&	79.5	&	0.029	\\
\addlinespace[2pt]
\hline \hline
    \end{tabular}
    \caption{Changes in the lowest vibrational frequency, $\Delta E_{vib}$ (cm$^{-1}$), and the 4/5$s$ Mulliken charge, $\Delta$Charge, compared between the large and the small active spaces for the hydrides and oxides.}
    \label{tab:Vib_E}
\end{table}
The differences in predicted frequencies between the large and small active spaces are generally smaller for the $4d$ transition metals, with the exception of NbH. CrH has a major improvement in the calculated vibrational frequency with the large active space, although this is unlikely to be caused solely by the second $d$-shell, as we found this example to be quite sensitive to the CASPT2 parameters. The vibrational frequencies of the oxides with the larger active space are generally in better agreement with known experimental values, while the hydrides have a similar accuracy for both active spaces, as shown in Supporting Information Table S6.~\cite{Armentrout:1996aa,Scullman:1975aa,Wang:2013aa,Friedman-Hill:1992aa,Namiki:1996aa,Chestakov:2005aa,Cheng:2017aa,Bauschlicher:2001aa,Barnes:1997aa,Ram:1993aa,Cheung:1982aa} This level of theory is likely insufficient for quantitative results, which would require larger basis sets and more sophisticated treatment of relativistic effects; however, it provides a good qualitative picture of how the second $d$-shell effect is important for simulating quantities which are sensitive to the dynamical correlation.\cite{Aoto:2017aa,Cheng:2017aa,Balabanov:2006aa,Phung:2018aa,Sorensen:2020aa,Vavrecka:aa}

\begin{figure*}[ht!]
\includegraphics[scale = 0.45]{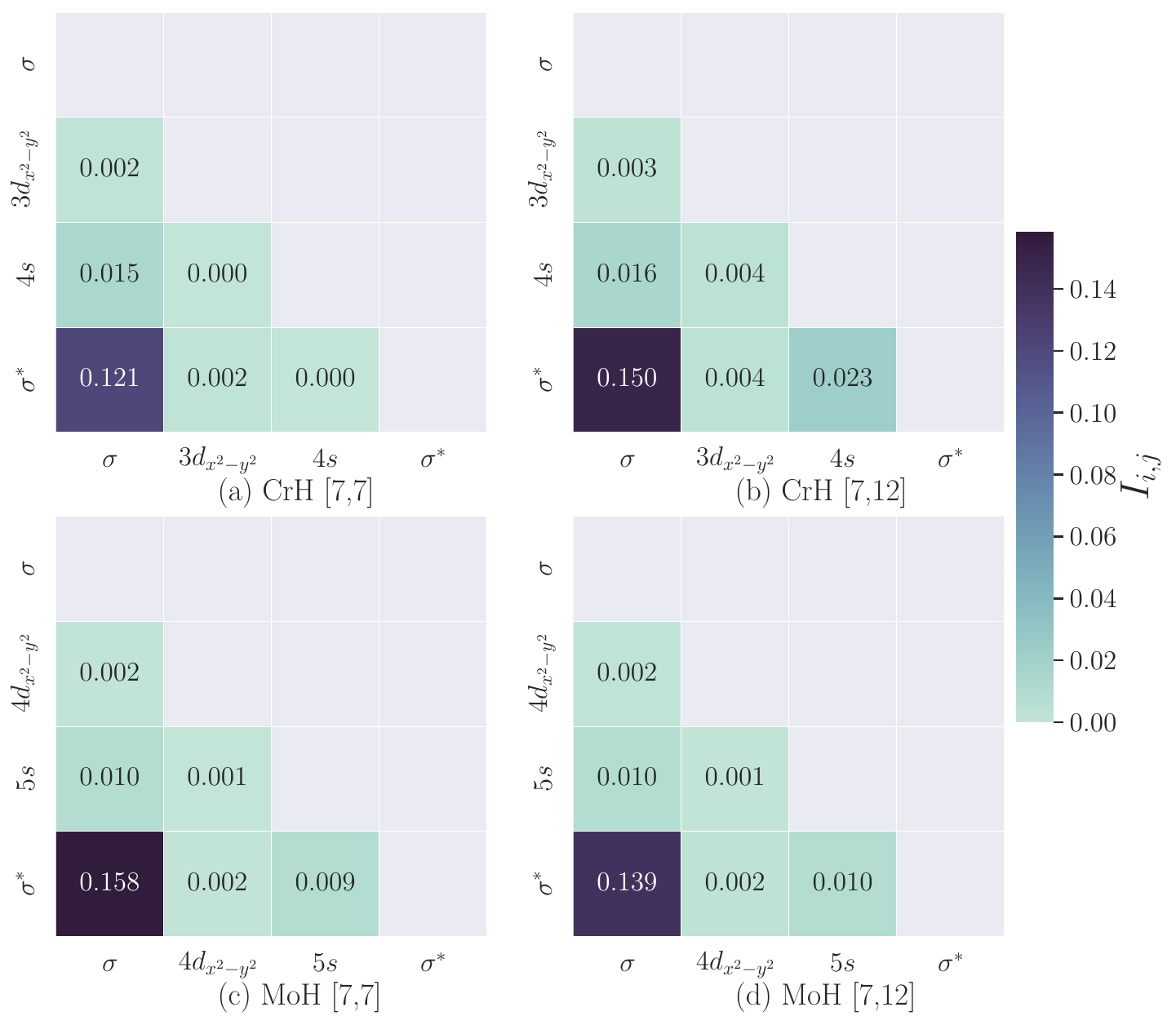}
\caption{Mutual information ($I_{ij}$) heatmaps for CrH with the small (a) and large (b) active spaces, and heatmaps for MoH with the small (c) and large (d) active spaces, showing a subset of orbitals of $A_1$ symmetry. The diagonal elements are zero, and the mutual information is symmetric about the diagonal. The maximum entropy between these orbital pairs is about 0.16.}
\label{fig:CrMo_heatmap}
\end{figure*}

Another important signature of the second $d$-shell effect is the population or Mulliken charge of the 4$s$ orbital, which tends to be overestimated when the second $d$-shell effect is prominent,~\cite{Pierloot:2001aa,Dunning:1980aa} shown in Table~\ref{tab:Vib_E}. For the hydrides, the $4/5s$ charge remains unchanged when increasing the size of the active space, which is expected for the early transition metals.~\cite{Pierloot:1995aa} On the other hand, the $3d$ oxides exhibit significant changes in the Mulliken charges when including the second $d$-shell, denoted as the $d'$ orbitals. This is consistent with the one-orbital entropy metrics for the 3$d$ orbitals, which tend to be underestimated by the smaller active space, shown in Tables~\ref{tab:ooeH} and ~\ref{tab:ooeO} for the hydrides and oxides respectively.
\begin{table}[ht!]
    \centering
    \begin{tabular}{@{\extracolsep{4pt}}l c c c c c c}
    \hline \hline
    & \multicolumn{6}{c}{Small Active Space} \\
    \cline{2-7}
&	TiH	&	VH	&	CrH	&	ZrH	&	NbH	&	MoH	\\
\cline{2-2}\cline{3-3}\cline{4-4}\cline{5-5}\cline{6-6}\cline{7-7}
$\sigma$	&	0.07	&	0.07	&	0.14	&	0.07	&	0.11	&	0.17	\\
$d_{x^2-y^2}$	&	0.69	&	0.00	&	0.01	&	0.69	&	0.01	&	0.00	\\
$\sigma^*$	&	0.07	&	0.07	&	0.14	&	0.07	&	0.10	&	0.17	\\
$4/5s$	&	0.02	&	0.02	&	0.03	&	0.01	&	0.01	&	0.02	\\
$d_{xz}$	&	0.69	&	0.00	&	0.00	&	0.70	&	0.01	&	0.01	\\
$d_{xy}$	&	0.69	&	0.00	&	0.00	&	0.69	&	0.00	&	0.00	\\
$d_{yz}$	&	0.69	&	0.00	&	0.00	&	0.70	&	0.01	&	0.01	\\
\hline
& \multicolumn{6}{c}{Large Active Space} \\
    \cline{2-7}
&	TiH	&	VH	&	CrH	&	ZrH	&	NbH	&	MoH	\\
\cline{2-2}\cline{3-3}\cline{4-4}\cline{5-5}\cline{6-6}\cline{7-7}
$\sigma$	&	0.16	&	0.16	&	0.21	&	0.15	&	0.17	&	0.22	\\
$d_{x^2-y^2}$	&	0.69	&	0.05	&	0.06	&	0.69	&	0.01	&	0.03	\\
$\sigma^*$	&	0.10	&	0.14	&	0.23	&	0.09	&	0.12	&	0.19	\\
$4/5s$	&	0.06	&	0.06	&	0.09	&	0.05	&	0.02	&	0.04	\\
$d^\prime_{x^2-y^2}$	&	0.03	&	0.04	&	0.05	&	0.02	&	0.12	&	0.03	\\
$d^\prime_{z^2}$	&	0.02	&	0.03	&	0.03	&	0.03	&	0.01	&	0.03	\\
$d_{xz}$	&	0.70	&	0.06	&	0.06	&	0.70	&	0.04	&	0.04	\\
$d^\prime_{xz}$	&	0.02	&	0.05	&	0.05	&	0.02	&	0.04	&	0.04	\\
$d_{xy}$	&	0.69	&	0.01	&	0.06	&	0.69	&	0.01	&	0.03	\\
$d^\prime_{xy}$	&	0.02	&	0.00	&	0.05	&	0.02	&	0.02	&	0.03	\\
$d_{yz}$	&	0.70	&	0.06	&	0.06	&	0.70	&	0.04	&	0.04	\\
$d^\prime_{yz}$	&	0.02	&	0.05	&	0.05	&	0.02	&	0.04	&	0.04	\\
\hline \hline
    \end{tabular}
    \caption{One-orbital entropies ($s_i$) for the valence and $d^\prime$ orbitals for the transition metal hydrides.}
    \label{tab:ooeH}
\end{table}
\begin{table}[ht!]
    \centering
    \begin{tabular}{@{\extracolsep{4pt}}l c c c c c c}
    \hline \hline
    & \multicolumn{6}{c}{Small Active Space} \\
        \cline{2-7}
	&	FeO	&	CoO	&	NiO	&	RuO	&	RhO	&	PdO	\\
\cline{2-2}\cline{3-3}\cline{4-4}\cline{5-5}\cline{6-6}\cline{7-7}
$\sigma$	&	0.53	&	0.39	&	0.42	&	0.35	&	0.25	&	0.20	\\
$d_{x^2-y^2}$	&	0.05	&	0.06	&	0.12	&	0.03	&	0.04	&	0.01	\\
$\sigma^*$	&	0.70	&	0.70	&	0.71	&	0.35	&	0.31	&	0.24	\\
$4/5s$	&	0.31	&	0.48	&	0.43	&	0.05	&	0.13	&	0.07	\\
$2p_x$	&	0.25	&	0.31	&	0.33	&	0.18	&	0.29	&	0.32	\\
$d_{xz}$	&	0.26	&	0.33	&	0.39	&	0.18	&	0.30	&	0.32	\\
$d_{xy}$	&	0.05	&	0.09	&	0.12	&	0.02	&	0.07	&	0.01	\\
$2p_y$	&	0.25	&	0.31	&	0.33	&	0.18	&	0.29	&	0.32	\\
$d_{yz}$	&	0.26	&	0.33	&	0.39	&	0.18	&	0.30	&	0.32	\\
\hline
& \multicolumn{6}{c}{Large Active Space} \\
    \cline{2-7}
	&	FeO	&	CoO	&	NiO	&	RuO	&	RhO	&	PdO	\\
\cline{2-2}\cline{3-3}\cline{4-4}\cline{5-5}\cline{6-6}\cline{7-7}
$\sigma$	&	0.40	&	0.30	&	0.25	&	0.30	&	0.22	&	0.20	\\
$d_{x^2-y^2}$	&	0.12	&	0.13	&	0.12	&	0.09	&	0.08	&	0.07	\\
$\sigma^*$	&	0.38	&	0.38	&	0.34	&	0.29	&	0.28	&	0.25	\\
$4/5s$	&	0.19	&	0.20	&	0.16	&	0.09	&	0.16	&	0.12	\\
$d^\prime_{x^2-y^2}$	&	0.15	&	0.10	&	0.10	&	0.07	&	0.07	&	0.06	\\
$d^\prime_{z^2}$	&	0.05	&	0.06	&	0.06	&	0.03	&	0.05	&	0.06	\\
$2p_x$	&	0.24	&	0.31	&	0.33	&	0.23	&	0.30	&	0.29	\\
$d_{xz}$	&	0.19	&	0.28	&	0.35	&	0.20	&	0.29	&	0.30	\\
$d^\prime_{xz}$	&	0.06	&	0.06	&	0.08	&	0.05	&	0.04	&	0.10	\\
$d_{xy}$	&	0.07	&	0.11	&	0.12	&	0.05	&	0.09	&	0.07	\\
$d^\prime_{xy}$	&	0.04	&	0.05	&	0.10	&	0.03	&	0.03	&	0.06	\\
$2p_y$	&	0.24	&	0.31	&	0.33	&	0.23	&	0.30	&	0.29	\\
$d_{yz}$	&	0.19	&	0.28	&	0.35	&	0.20	&	0.29	&	0.30	\\
$d^\prime_{yz}$	&	0.06	&	0.06	&	0.08	&	0.05	&	0.04	&	0.10	\\
\hline \hline
    \end{tabular}
    \caption{One-orbital entropies ($s_i$) for the valence and $d^\prime$ orbitals for the transition metal oxides.}
    \label{tab:ooeO}
\end{table}
Without the second $d$-shell, correlation is not properly accounted for, and preferential occupation of the $d^{N-1}s^1$ state is observed. Similar to the hydrides, the change in the Mulliken charge upon inclusion of the $d'$ orbitals is again negligible when considering the $4d$ oxides. 

These results highlight the importance of the $d'$ orbitals in calculating vibrational frequencies and Mulliken populations in the late third row transition metals, and they corroborate previous results reporting the practical consequences of the double-$d$ shell effect. Additionally, we use the entanglement entropy to reveal how the $d'$ orbitals subtly change important orbital interactions. These interactions can modulate both static and dynamic correlation effects, and their accurate description is paramount for accurate prediction of wavefunction properties.

For example, Figure~\ref{fig:CrMo_heatmap} shows the mutual information heatmap for CrH (top) and MoH (bottom) in the small (left) and large (right) active spaces. We show the mutual information between the valence orbitals of $A_1$ symmetry, which includes the $\sigma$, $\sigma^*$, $nd_{x^2-y^2}$, and $(n+1)s$ orbitals. The $\sigma$-$\sigma^*$ interaction is the most important among the orbitals shown, and the strength of entanglement remains the same order of magnitude with the addition of the $d'$ orbitals. The large mutual information for the $\sigma$-$\sigma^*$ interaction indicates that it is critical for the non-dynamical correlation required for the correct description of bond dissociation, as expected for the bonding orbitals; however, the one-orbital entropy also indicates important static correlation effects.~\cite{Boguslawski:2012aa,Suss:2020aa,Ding:2020aa} MoH has basically the same orbital entanglement interactions for both active spaces. On the other hand, CrH shows more small and moderate mutual information with the larger active space, which introduces more dynamic correlation into the wavefunction.

\begin{figure}[ht!]
    \centering
    \includegraphics[scale=0.5]{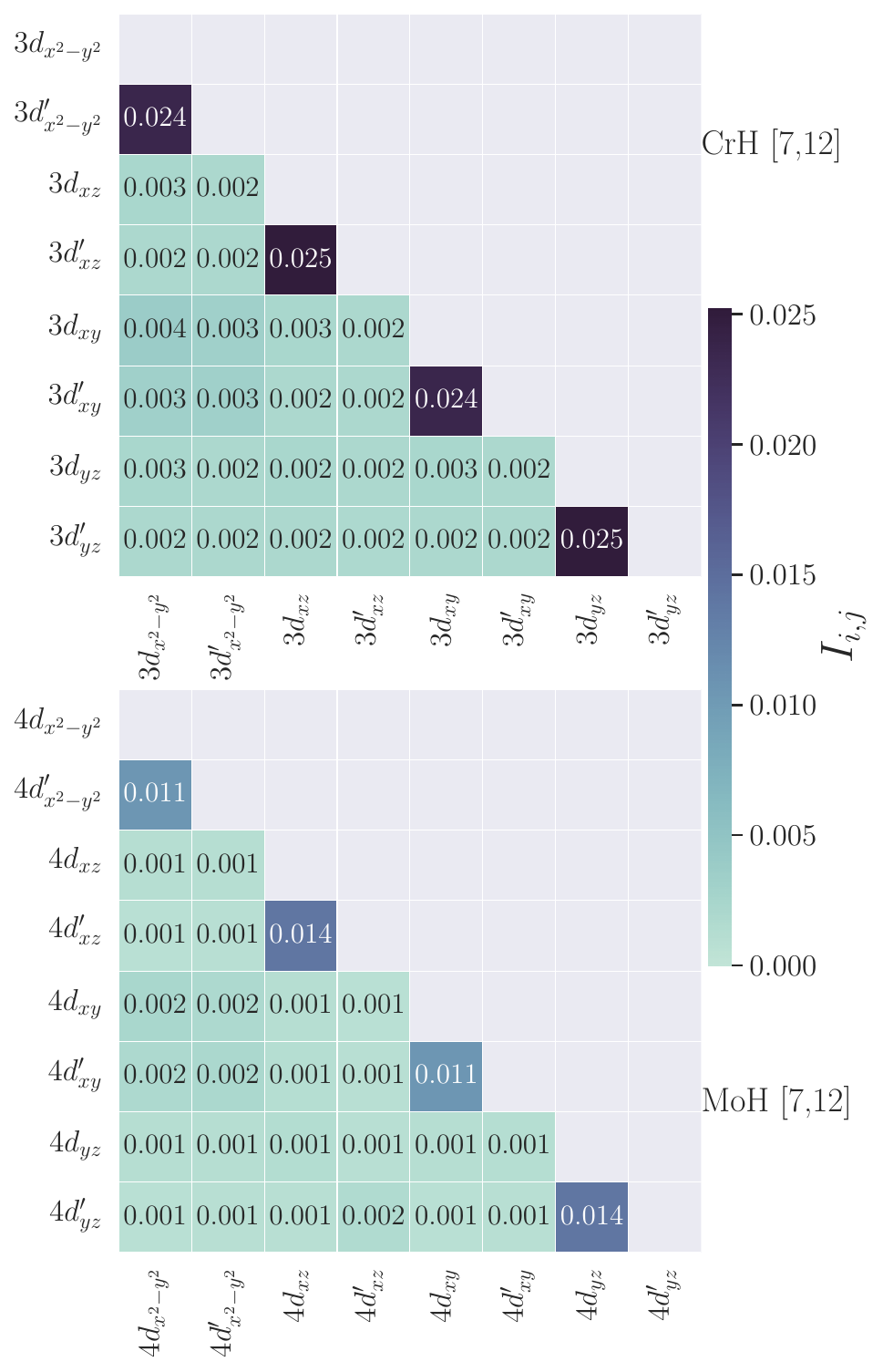}
    \caption{Mutual information ($I_{i,j}$) heatmaps for CrH (top) and MoH (bottom) with large active spaces, showing only the non-bonding $d$ orbitals. The maximum entropy between these orbital pairs is about 0.025.}
    \label{fig:CrH_dd_mutinf}
\end{figure}

The importance of the $d$-$d'$ interactions for the early transition metals can be further seen in Figure~\ref{fig:CrH_dd_mutinf}, which shows the $d$-$d'$ interactions in both CrH and MoH.
Among the $d$-$d'$ pairs, the MoH orbital entanglement is slightly smaller than CrH, but still relevant despite the addition of the $4d'$ orbitals into the active space. The magnitude of these interactions, along with the single orbital entropies (Table ~\ref{tab:ooeH}), indicate that these effects will be important for the dynamic correlation. This analysis is further verified by a DMRG-full configuration interaction (FCI) reference calculation for CrH with a smaller basis.~\cite{Roos2004,Roos2005,widmark1990a} When the entire orbital space is included in the FCI calculation, the mutual information between the $3d$-$3d'$ orbitals reduces. Thus the interaction between the $3d$-$3d'$ orbitals is likely not related to nondynamic correlation, because the mutual information and single orbital entropies of the $3d'$ orbitals are generally small. Nonetheless, the entanglement entropy indicates that the second $d$-shell interactions can mediate both non-dynamic and dynamic correlation effects. A more detailed discussion of the DMRG-FCI calculation is available in Supporting Information Section S4. 

\begin{figure*}[ht!]
    \centering
    \includegraphics[scale=0.5]{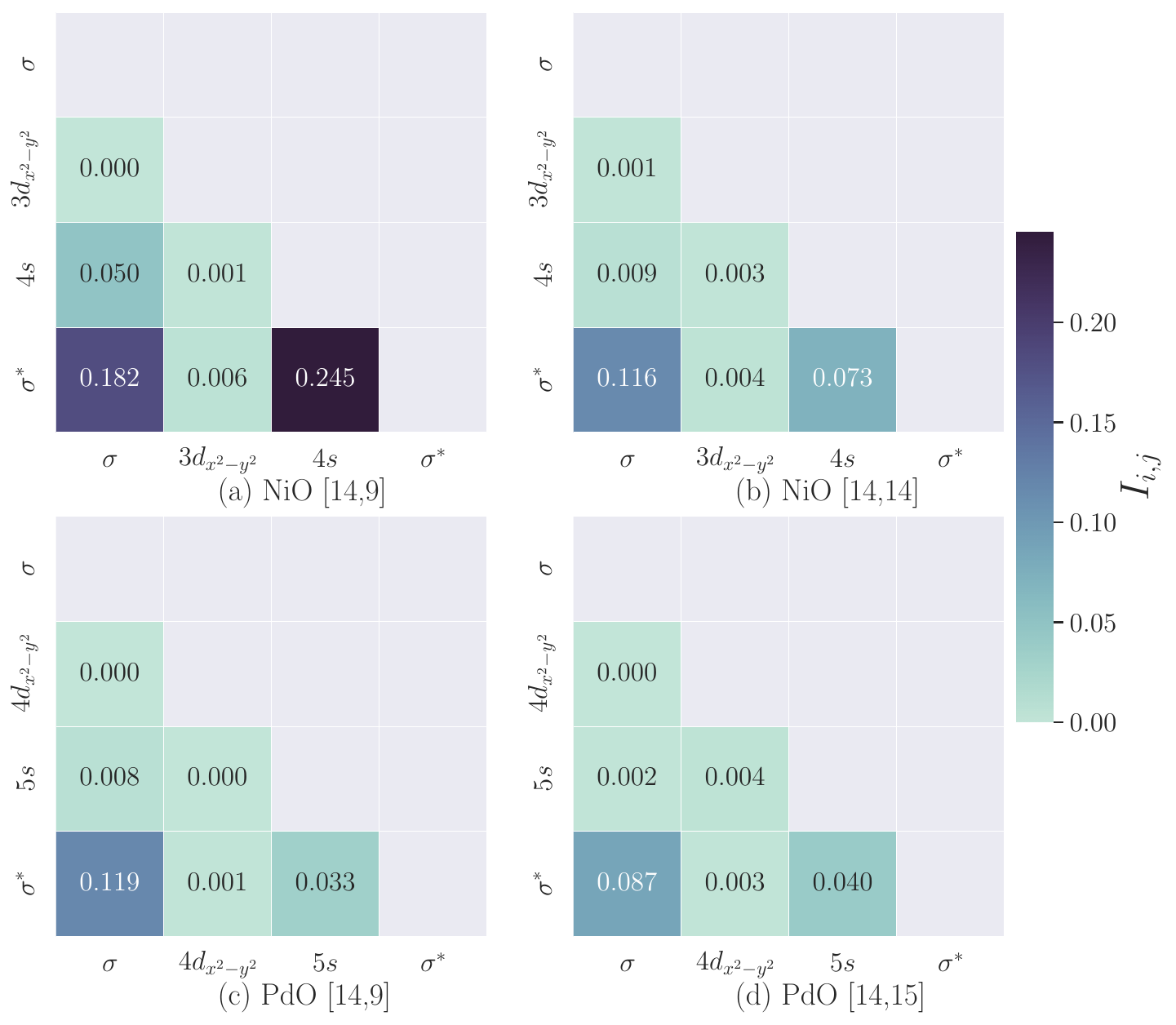}
    \caption{Mutual information ($I_{ij}$) heatmaps for NiO with the small (a) and large (b) active spaces, and heatmaps for PdO with the small (c) and large (d) active spaces, showing a subset of orbitals of $A_1$ symmetry. The maximum entropy between these orbital pairs is about 0.25.}
    \label{fig:NiO_PdO_heatmap}
\end{figure*}

The late 3$d$ transition metal oxides yield a starker contrast between the small and large active spaces. The wavefunction corresponding to the small active space indicates very strong interactions between the orbitals along the bonding axis, the $4s$ and $\sigma$-$\sigma^*$; however, the larger active space and the $4d$ metals suggest the smaller calculation overestimates the correlation. This is shown in Figure~\ref{fig:NiO_PdO_heatmap}, where we compare the mutual information for NiO and PdO using both active spaces.
Most significantly, NiO with only the $3d$ orbitals substantially overestimates the entanglement between the $4s$ and $\sigma^*$ orbitals, and the effect is much less significant in PdO. The mutual information between these two orbitals reduces from 0.25 to 0.07 when including the second $d$-shell. The one-orbital entropy also decreases for the $4s$ orbital, though it is still important for multi-reference correlation in the larger active space. This is also true of FeO, CoO and their $4d$ analogues, shown in Supporting Information Sections S5-S6, and examples of this over-correlation have been reported in other cases as well.\cite{Boguslawski:2012aa}

The over-correlation of the $4s$ orbital is somewhat paradigmatic of the double $d$-shell effect,~\cite{Dunning:1980aa} but the analysis of the orbital entanglement indicates a somewhat complex picture. In this example, the $4s$ orbital is clearly important for multi-reference correlation, but the small active space provides a completely unbalanced view of the total correlation. An analysis of the important configurations for these wavefunctions supports this, shown in Supporting Information Table S7, indicating that both static and dynamic correlation are poorly described by the smaller active space.

\section{Discussion and Conclusions}

Here, we characterize the second $d$-shell effect in a periodic series of 3$d$ and 4$d$ transition metal hydride and oxides by considering the quantum information entropy of correlated wavefunctions. For these molecular series, we present changes in the lowest vibrational frequency, one-orbital entropies, and mutual information heatmaps for relevant molecular orbitals. These results highlight the importance of the second $d$-shell in modulating subtle electronic interaction in molecular transition metal ions. Much of the utility of these species derives from the rich electronic structure provided by the $d$ manifold; however, calculating a balanced wavefunction that incorporates various correlation effects is highly non-trivial. Perhaps the most significant barrier to accurate calculation of properties arises in dynamical correlation corrections, which are heavily influenced by the underlying multi-reference wavefunction. Noteably, many of the significant consequences of the second $d$-shell effect can be seen in this series diatomic molecules, which have relatively simple metal-ligand interactions.

Similar behavior also occurs in more complicated transition metal systems where spin delocalization can be important. For example copper dithiolate systems, which are seen as promising molecular qubit candidates,~\cite{Bader:2014aa,Kazmierczak:2021aa,Fataftah:2019aa} have important interactions due to the Cu-S bonds, and previous work has demonstrated the importance of the second $d$-shell in modeling the spin populations of Cu and S. Experiments show that the unpaired spin in the system is delocalized, with about 75\% of the spin density on Cu. Calculations corroborate the experiment when the second $d$-shell is included, but the valence calculation predicts the spin is totally localized on Cu.~\cite{Fataftah:2019aa,Schlimgen:2023aa}

Here we have shown that the second $d$-shell effect can be characterized using orbital entanglement entropy derived from correlated wavefunctions. Some of these results are quite expected: the second $d$-shell effect is most important for late transition metals in the sense that the electronic density can be qualitatively incorrect, as demonstracted by the Mulliken populations of the late oxides, for example. Perhaps contrary to expectations, the early transition metals are also sensitive to the effect. The entanglement entropy analysis showed that in CrH, for example, the second $d$-shell plays an important role in mediating static and dynamic correlation. Finally, we computed an approximate FCI wavefunction for CrH that shows the second $d$-shell interactions are in an ambiguous regime of static and dynamic correlation.

For active space methods, one of the most important considerations for quantitatively accurate results is balancing the reference wavefunction such that the dynamic and static correlation perturbatively approximate the exact wavefunction, typically using CASPT2 or $N$-electron-valence perturbation theory. Contemporary applications using transition metals, especially for quantum computing and materials science, often require highly accurate descriptions of excited states and spectroscopic properties. Our results provide important insights into how the second $d$-shell effect complicates the influence of electron correlation in different transition metal bonding environments. Another important consideration is orbital localization, which is known to affect the picture of correlation in quantum systems.~\cite{Olivares-Amaya:2015aa,Ratini:2024aa,Szalay:2017aa,Ding:2021aa,Ding:2023aa,Ding:2024aa} Localization and approaches utilizing multiple $d$-occupancy wavefunctions~\cite{Moraes:2023aa} could be combined with the present entanglement entropy analysis to expose further the complicated correlateld electronic structure in transition metal chemistry.

\textcolor{black}{Because active spaces are typically constructed by chemical intution, or informed by metrics like mutual information used here, one might be tempted to draw universal conclusions about inclusion of the second $d$-shell. In fact, our results demonstrate that these orbital interactions highly depend on the electronic configuration of the metal (early or late transiton metal) and the bonding environment, and naively adding these orbitals to the active space is unlikely to provide consistent results. This is largely due to the non-variational nature of multireference perturbation theory combined with the multi-faceted physical interactions made possible by $d$-orbital molecular bonding. The second $d$-shell can modulate static, dynamic, and/or non-dynamic correlation, and inclusion of particular orbitals in the active space should be analyzed on a case-by-case basis when accompanied by a perturbative treatment of the dynamic correlation. These issues are likely also important when considering excited state transitions, because of the changing electronic state of the species considered. Our results indicate that simple solutions to the second $d$-shell problem likely are not forthcoming in the context of multireference perturbation theory.}

The approaches to understand the interplay of static and dynamic correlation in transition metal complexes described here should be useful in practical construction of correlated electronic structure calculations. These techniques can be used to understand the shifting roles of static and dynamic correlation in correlated wavefunctions describing transition metals in problem-specific contexts. While many aspects of active space selection can be automated,~\cite{Lei:2021aa,Stein:2016aa,Zou:2020aa,Bao:2018aa,Bao:2019aa,Agarawal:2024aa} many advanced applications of transition metal chemistry will still require problem specific choices of relevant electronic states and orbitals. Informing these choices with mutual information analysis provides an avenue for understanding electron correlation in transition metal chemistry from a rigorous information entropy perspective.

\section{Supplementary Material}
Active space compositions, computed internuclear distances, vibrational frequencies, details of the DMRG-FCI calculation, remaining mutual information heatmaps, and configuration weights are availabe in the Supporting Information (PDF). 

\section{Acknowledgements}

KHM acknowledges the start-up funds from Washington University in St. Louis and the University of Minnesota. 

\section{Data Availability Statement}
The data that support the findings of this study are available from the corresponding authors upon reasonable request.

\renewcommand*{\bibfont}{\normalsize}
\bibliography{main}

\end{document}